\documentclass[letters,fleqn,useAMS,usenatbib]{mnras}
\usepackage{psfrag,color,epsfig}

\usepackage{graphicx}	
\usepackage{amsmath}	
\usepackage{amssymb}	
\usepackage{multicol}	
\usepackage{bm}		
\usepackage{pdflscape}	
\usepackage{multirow}
\usepackage{tabularx}  
\usepackage{caption}
\usepackage{newtxtext,newtxmath}

\usepackage[T1]{fontenc}
\usepackage{ae,aecompl}

\def\HI{\ion{H}{I}\,}

\def\xb{\bar{x}_{\rm \ion{H}{I}}}
\def\Tb{{T_{\rm b}}}
\def\tTb{\tilde{T}_{\rm b}}
\def\k{{\bm{k}}}

\def\x{{\bm{x}}}

\def\U{{\bm{U}}}

\def\cl{{\mathcal C}_{\ell}}
\def\dl{{\mathcal D}_{\ell}}

\usepackage{tikz,xcolor,hyperref}
\definecolor{lime}{HTML}{A6CE39}
\DeclareRobustCommand{\orcidicon}{%
	\begin{tikzpicture}
	\draw[lime, fill=lime] (0,0) 
	circle [radius=0.16] 
	node[white] {{\fontfamily{qag}\selectfont \tiny ID}};
	\draw[white, fill=white] (-0.0625,0.095) 
	circle [radius=0.007];
	\end{tikzpicture}
	\hspace{-2mm}
}
\foreach \x in {A, ..., Z}{%
	\expandafter\xdef\csname orcid\x\endcsname{\noexpand\href{https://orcid.org/\csname orcidauthor\x\endcsname}{\noexpand\orcidicon}}
}

\title[The MAPS in parameter studies]{The multi-frequency angular power spectrum in parameter studies of the cosmic 21-cm signal}
\author[R. Mondal et al.]{Rajesh Mondal\orcidA,$^{1}$\thanks{E-mail: rajesh@astro.su.se} Garrelt Mellema\orcidB,$^1$ Steven G. Murray\orcidC$^{2}$ and Bradley Greig\orcidD$^{3,4}$\\
$^{1}$The Oskar Klein Centre, Department of Astronomy, Stockholm University, AlbaNova, SE-10691 Stockholm, Sweden \\
$^{2}$School of Earth and Space Exploration, Arizona State University, Tempe, AZ, USA \\
$^{3}$School of Physics, University of Melbourne, Parkville, VIC 3010, Australia \\
$^{4}$ARC Centre of Excellence for All-Sky Astrophysics in 3 Dimensions (ASTRO 3D) \\
}

\date{Accepted 2022 May 10. Received 2022 May 04; in original form 2022 March 21}
\pubyear{2022}

\begin{document}
\label{firstpage}
\pagerange{\pageref{firstpage}--\pageref{lastpage}}
\maketitle


\begin{abstract}
The light-cone effect breaks the periodicity and statistical homogeneity (ergodicity) along the line-of-sight direction of cosmological emission/absorption line surveys. The spherically averaged power spectrum\,(SAPS), which by definition assumes ergodicity and periodicity in all directions, can only quantify some of the second-order statistical information in the 3D light-cone signals and therefore gives a biased estimate of the true statistics. The multi-frequency angular power spectrum\,(MAPS), by extracting more information from the data, does not rely on these assumptions. It is therefore better aligned with the properties of the signal. We have compared the performance of the MAPS and SAPS metrics for parameter estimation for a mock 3D light-cone observation of the 21-cm signal from the Epoch of Reionization. Our investigation is based on a simplified 3-parameter {\sf 21cmFAST} model. We find that the MAPS produces parameter constraints which are a factor of $\sim 2$ more stringent than when the SAPS is used. The significance of this result does not change much even in the presence of instrumental noise expected for 128\,hours of SKA-Low observations. Our results therefore suggest that a parameter estimation framework based on the MAPS metric would yield superior results over one using the SAPS metric.
\end{abstract}

\begin{keywords}
methods: statistical – techniques: interferometric – dark ages, reionization, first stars – large-scale structure of Universe – cosmology: observations – cosmology: theory. 
\end{keywords}


\section{Introduction}
\label{sec:intro}
The 21-cm signal produced by neutral hydrogen (\ion{H}{I}) in the Intergalactic Medium (IGM) during the Epoch of Reionization (EoR) encodes the answer to several key questions about reionization. Significant efforts are underway to measure the 21-cm Spherically Averaged Power Spectrum\,(SAPS) by ongoing and upcoming radio interferometric experiments, which can measure the signal in the sky at different frequencies and thus in three dimensions. Examples of such experiments are LOFAR\footnote{\url{http://www.lofar.org}} \citep{LOFAR-EoR:2020}, MWA\footnote{\url{http://www.haystack.mit.edu/ast/arrays/mwa}} \citep{Trott2020}, 
GMRT\footnote{\url{http://www.gmrt.ncra.tifr.res.in}} \citep{Paciga:2013}, HERA\footnote{\url{http://reionization.org}} \citep{TheHERACollaboration2021} and the future SKA\footnote{\url{http://www.skatelescope.org}} \citep{mellema13}.
Although for now, the primary goal remains the first detection, eventually it will be the measurement of the evolution of the SAPS at a range of spatial scales that will be crucial for our understanding of the EoR; for example, determining key IGM properties and astrophysical parameters such as the average ionization fraction, bubble size distributions, average ionizing emissivities and characteristic masses of the galaxies responsible for reionization \citep[e.g.][]{ghara2020, Mondal2020a, Greig2021a, Greig2021, 2022ApJ...924...51A}.

It has long been known that the SAPS has some intrinsic drawbacks as a summary statistic for the 21-cm signal. The strongly non-Gaussian character of the 21-cm signal has for example prompted the exploration of the bispectrum as an additional summary statistic \citep[e.g.][]{bharadwaj05a, Watkinson2017, 2020MNRAS.499.5090M, Mondal2021}. Another complication for the SAPS is caused by the so-called light-cone\,(LC) effect \citep{barkana06, datta12}, i.e. the evolution of the signal along the frequency direction, which breaks the statistical homogeneity (or ergodicity) along the line-of-sight (LoS) direction \citep{mondal18}. The effect is particularly important for 21-cm observations as the mean (global) brightness temperature changes rapidly as the universe evolves \citep{mondal2019}. \citet{mondal18} have shown that the SAPS can only quantify a part of the entire second-order statistics of the 21-cm signal and gives a biased estimate of the true two-point statistics \citep{trott16}, as it assumes the signal is ergodic and periodic in all three directions. One can reduce these effects by analysing the signal over small frequency intervals. However, this reduces accuracy as we miss out on the large-scale LoS modes. We therefore use a different metric, the Multi-frequency Angular Power Spectrum\,(MAPS), which does not rely on these assumptions \citep{Santos2005, datta07a, mondal18}. An additional benefit is that the MAPS is much closer to the visibility correlations as measured by the radio interferometers.

Constraining reionization models in terms of their model parameters has become possible due to the development of fast codes for simulating reionization and an increase in computing power. Therefore, the choice of an appropriate metric is of fundamental importance. Interpreting the data with an inaccurate metric will reduce the accuracy of the constraints on the model parameters. As a proof of concept, in this {\it letter}, we compare the performance of SAPS and MAPS within a parameter estimation framework. This is a pure statistical comparison and therefore the results and analyses presented here are generic and valid for the statistical analysis of any 3D signal which is non-ergodic and non-periodic along one of its axes.


Throughout the {\it letter}, we have used the Planck+WP best fit values of cosmological parameters \citep[][table 2, last column]{PlanckCollaboration2020}.

\section{Simulating the light-cone 21-cm signal}
\label{sec:sim}
To simulate 21-cm light-cones we use the publicly available semi-numerical code {\sf 21cmFAST}\footnote{\url{https://github.com/21cmFAST/21cmFAST}} \citep{Mesinger:2011, Murray2020}. In particular, we simulate light-cones between $z=7.21$ to 8.90, and assume the spin temperature is saturated ($T_\mathrm{S} \gg T_\mathrm{CMB}$, i.e. we do not model spin temperature fluctuations). 

Specifically, we use the simple 3-parameter astrophysical model from \citet{Greig2015}, foregoing the complexity of newer models that incorporate e.g. PopIII stars, and we also ignore redshift-space distortions and inhomogeneous recombinations. These three parameters are: the ionizing efficiency ($\zeta$), the mean free path of ionizing photons within ionizing regions ($R_{\rm mfp}$) and the minimum virial temperature of star-forming halos ($T_{\rm vir}$). These have been shown to span the greater portion of reasonable physical models \citep{Greig2015}, and are adequate for our proof-of-concept work here. We adopt  $[\zeta,\,R_{\rm mfp},\,\log(T_{\rm vir}) ] = [34,\,15,\,\log{(50000)}]$ as our fiducial values. For our fiducial model, the change in the \HI fraction, which characterizes the strength of the LC effect, is $\Delta \xb = 0.68-0.26 = 0.42$ over the aforementioned $z$ range.

We perform simulations in volumes of size $[500\,{\rm Mpc}]^3$, in which the initial density field is computed on a $N=512^3$ grid, which is smoothed down to an $N=128^3$ grid on which the astrophysics is simulated. 



\begin{figure*}
\centering
\psfrag{nu2}[c][c][1.2]{$\nu_2 - \nu_{\rm c}$ (MHz)}
\psfrag{nu1}[c][c][1.2]{$\nu_1 - \nu_{\rm c}$ (MHz)}
\psfrag{cl}[c][c][1.2]{$\dl(\nu_1, \nu_2)~({\rm mK})^2$}
\includegraphics[width=.95\textwidth, angle=0]{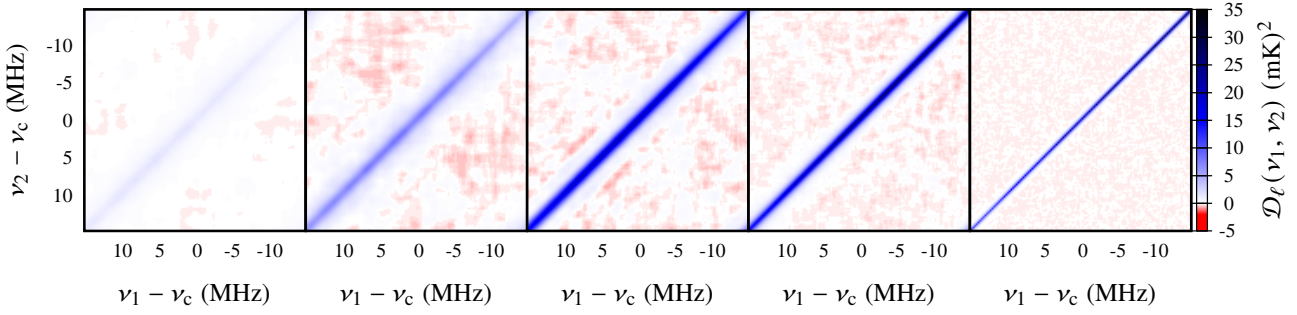}
\caption{The $\dl(\nu_1, \nu_2)$ at $\ell = 133.94$, 348.10, 763.50, 1774.1 and 4081.7 (from left to right respectively) for the fiducial model, where $\nu_{\rm c} = 158.2$\,MHz.}
\label{fig:cl_nu12}
\end{figure*}



\begin{figure}
\psfrag{cl}[c][c][1][0]{\large $\dl^{\rm EP}(\Delta
  \nu)/(2\pi)~{\rm mK}^2$}  
\psfrag{deltanu}[c][c][1][0]{$(\Delta \nu+1)$~{MHz}}
\psfrag{l=133.74}[c][c][1][0]{$\ell=133.74$}
\psfrag{348.10}[c][c][1][0]{~~\,$348.10$}
\psfrag{763.50}[c][c][1][0]{~~\,$763.50$}
\psfrag{1774.08}[c][c][1][0]{~~\,$1774.1$}
\psfrag{4081.71}[c][c][1][0]{~~\,$4081.7$}
\centering
\includegraphics[width=0.45\textwidth, angle=0]{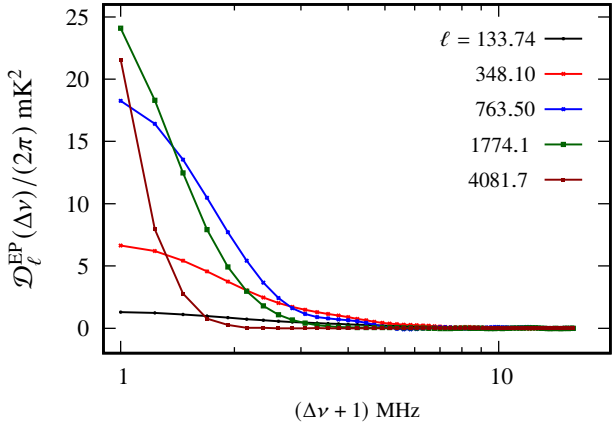}
\caption{The $\dl^{\rm EP}(\Delta \nu)$ as a function of $\Delta \nu$ for the $\ell$ bins considered in  Fig.\,\ref{fig:cl_nu12}. The $\Delta \nu$ values have been shown for half the bandwidth as the signal is periodic. Note we have shown $\Delta \nu+1$ rather than $\Delta \nu$ to avoid $\Delta \nu = 0$ points.
}
\label{fig:cl_deltanu}
\end{figure}



\begin{figure*}
\centering
\psfrag{nu1}[c][c][1.2]{$\nu - \nu_{\rm c}$ (MHz)}
\psfrag{cl}[c][c][1.2]{$\dl~({\rm mK})^2$}
\psfrag{cl-nu11}[c][c][1.2]{$\dl(\nu,\nu)$~~}
\psfrag{cl-dnu0}[c][c][1.2]{$\dl^{\rm EP}(0)$~~}
\psfrag{cl-nu13}[c][c][1.2]{$\dl(\nu,\nu+.5)$~~~~~~~~}
\psfrag{cl-dnu2}[c][c][1.2]{$\dl^{\rm EP}(0.5)$~~~~}
\includegraphics[width=.95\textwidth, angle=0]{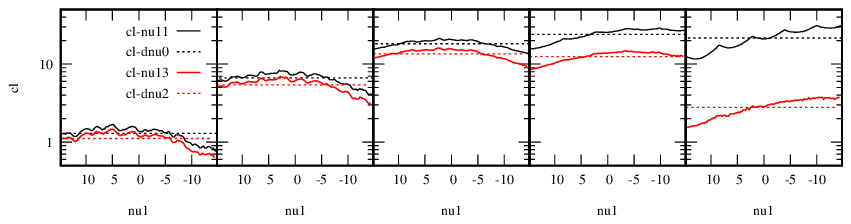}
\caption{This shows the diagonal (black solid lines) and 2nd off-diagonal (red solid lines) terms of MAPS i.e. $\dl(\nu, \nu)$ and $\dl(\nu, \nu+0.46)$ at the four different $\ell$ bins considered in  Fig.\,\ref{fig:cl_nu12}. We also show the $\dl^{\rm EP}(\Delta \nu)$'s correspond to the average of $\dl(\nu_1, \nu_2)$'s over all possible combinations of $\nu$ for frequency separations $\Delta \nu = 0$ and $\Delta \nu = 0.46$\,MHz, respectively.}
\label{fig:cl_nu11}
\end{figure*}


\section{Statistical Analysis Methodology}
\label{sec:method}
We define the MAPS using 
\begin{equation}
\cl(\nu_1,\,\nu_2) = {\mathcal C}_{2\pi{\rm U}}(\nu_1,\,\nu_2) =\frac{1}{\Omega} \big\langle \tTb(\U,\nu_1) \tTb(-\U,\nu_2) \big\rangle\,,
\label{eq:cl}
\end{equation}
where $\ell= 2\upi\U$ is the angular multipole, $\Omega$ is the solid angle subtended by the simulation at the observer, angle brackets represent an ensemble average and $\tTb(\U,\nu)$ is the 2D Fourier transform of the brightness temperature fluctuations $\delta \Tb(\theta,\nu)$. In this work, we have adopted the flat-sky approximation where $\theta$ is the 2D vector on the plane of the sky and its Fourier conjugate is denoted by $\U$. The definition of MAPS assumes that the signal is statistically homogeneous and isotropic on the plane of the sky but it does not assume this for the LoS direction.

As mentioned, the SAPS $P(\k)$ assumes the signal to be ergodic\,(E) and periodic\,(P) in all three directions. 
If we were to impose statistical homogeneity on $\cl(\nu_1,\,\nu_2)$ along the frequency direction, it is obvious that its values will only depend on the distance along the LoS, or in other words, on the frequency separation $\Delta \nu = |\nu_1 - \nu_2|$. If we further impose periodicity, we have $\cl^{\rm EP}(\Delta \nu) = \cl^{\rm EP}(B - \Delta \nu)$, where $B$ is the frequency bandwidth over which we are extracting the $\cl$. Under these two assumptions, we can write the relation between $P(\k)$ and $\cl^{\rm EP}(\Delta \nu)$ as (Eq.\,14 of \citealt{mondal18})
\begin{equation}
P(\k) = P(\k_{\perp},\,k_{\parallel})= r_{\rm c}^2\,r^{\prime}_{\rm c} \int d (\Delta \nu) \, e^{-i  k_{\parallel} r^{\prime}_{\rm c} \Delta  \nu}\, \cl^{\rm EP}(\Delta \nu)\,,
\label{eq:cl_Pk}
\end{equation}
where $\k_{\perp} = \ell/r_{\rm c}$ and $k_{\parallel}$ are the perpendicular and parallel components of $\k$ to the LoS, respectively. Here, $r_{\rm c}$ is the comoving distance to the centre of the light-cone and $r^{\prime}_{\rm c} = \frac{d r}{d \nu}|_{r_{\rm c}}$. Eq.\,\ref{eq:cl_Pk} shows that $P(\k)$ is essentially the Fourier transform of $\cl^{\rm EP}(\Delta \nu)$ along the frequency axis. Therefore below, we use $\cl^{\rm EP}(\Delta \nu)$ as a proxy for $P(\k)$ to keep MAPS and SAPS metrics in the same dimension.


%

We have used our publicly available {\sf MAPS}\footnote{\url{https://github.com/rajeshmondal18/MAPS}} code \citep{mondal18, mondal2019, mondal2020} to calculate the bin averaged $\cl(\nu_1,\,\nu_2)$ and $\cl^{\rm EP}(\Delta \nu)$. We have used 10 equally spaced logarithmic $\ell$ bins in the range $[114.74,\, 7342.9]$. We use the scale-independent $\dl(\nu_1, \nu_2) = \ell(\ell+1)\cl(\nu_1, \nu_2)/2\upi$ and $\dl^{\rm EP}(\Delta \nu) = \ell (\ell+1) \cl^{\rm EP}(\Delta \nu) / 2\upi$ in our analysis. Fig.\,\ref{fig:cl_nu12} shows the average $\dl(\nu_1, \nu_2)$ at $\ell = 133.94$, 348.10, 763.50, 1774.1 and 4081.7 (from left to right respectively) for the fiducial model using 100 independent realizations, where the central frequency $\nu_{\rm c} = 158.2$\,MHz. The two main features of MAPS are that its magnitude peaks along the diagonal and falls off rapidly away from it as the frequency separation increases. This is more clear in Fig.\,\ref{fig:cl_deltanu} where we show the corresponding $\dl^{\rm EP}(\Delta \nu)$. We see that the signal falls off faster for higher $\ell$ values but oscillates close to zero beyond $\Delta \nu \ga 4$\,MHz for all $\ell$.

To understand why the power spectrum gives a biased estimate of the true 2-point statistics and captures only the ergodic and periodic part of the information in the signal, we plot the diagonal and 2nd off-diagonal terms of the MAPS matrix as a function of $\nu$ in Fig.\,\ref{fig:cl_nu11}. It shows the variation of the 2-point statistics of the signal along the frequency direction. In Fig.\,\ref{fig:cl_nu11}, we also show the $\dl^{\rm EP}(\Delta \nu)$, which does not vary with $\nu$, corresponds to the average of $\dl(\nu_1, \nu_2)$ over all possible combinations of $\nu_1$ and $\nu_2$ which are $\Delta \nu$ apart. We find that $\dl(\nu_1, \nu_2)$ shows a systematic statistical variation along the $\nu$ direction as compared to $\dl^{\rm EP}(\Delta \nu)$, which signifies that the power spectrum misses out some 2-point statistical information in the signal. We do not discuss these further here and refer the reader to the discussion of fig.\,9 in \citet{mondal18}.

Fig.\,\ref{fig:cl_nu11} also shows that at small scales (right-hand panel) our MAPS results show an oscillatory pattern with $\nu$ which is not statistical in nature. The origin of this is likely a mix of noise on small-scales from the excursion set approach mixed with the stitching of boxes to construct the light-cones in {\sf 21cmFAST}. Simulated light-cone data from the {\sf C$^2$-ray} \citep{mellema06} and {\sf ReionYuga} \citep{mondal15, mondal17} do not show this spurious feature. How we deal with this is discussed in Sec.\,\ref{sec:error}.


\begin{figure*}
\centering
\includegraphics[width=.33\textwidth, angle=0]{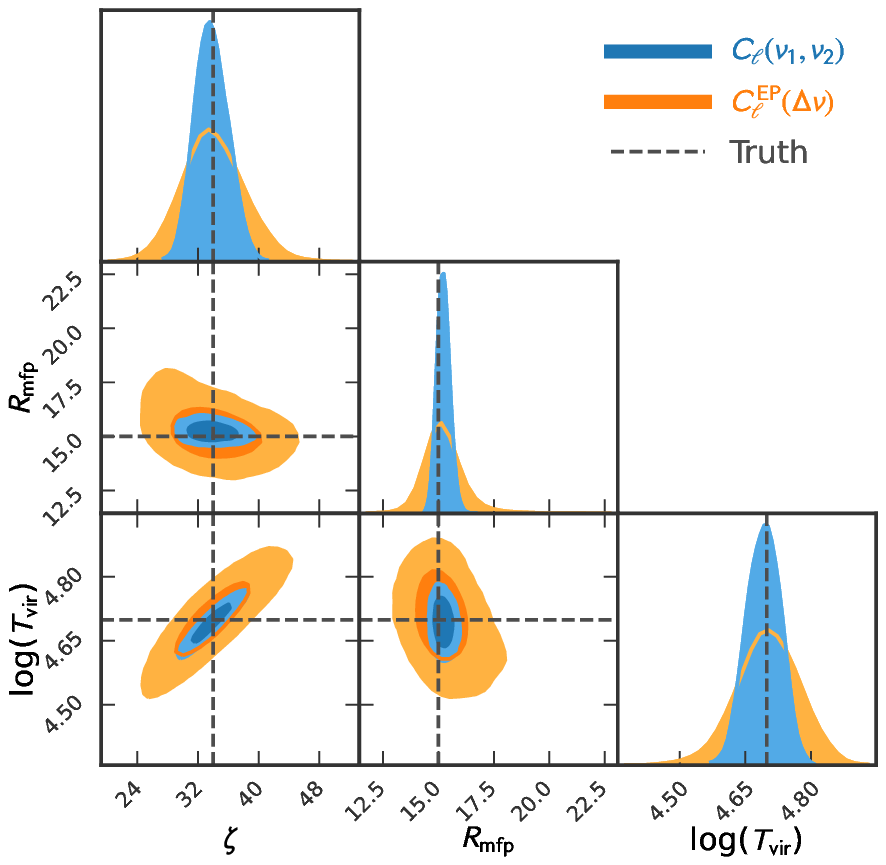}
\includegraphics[width=.33\textwidth, angle=0]{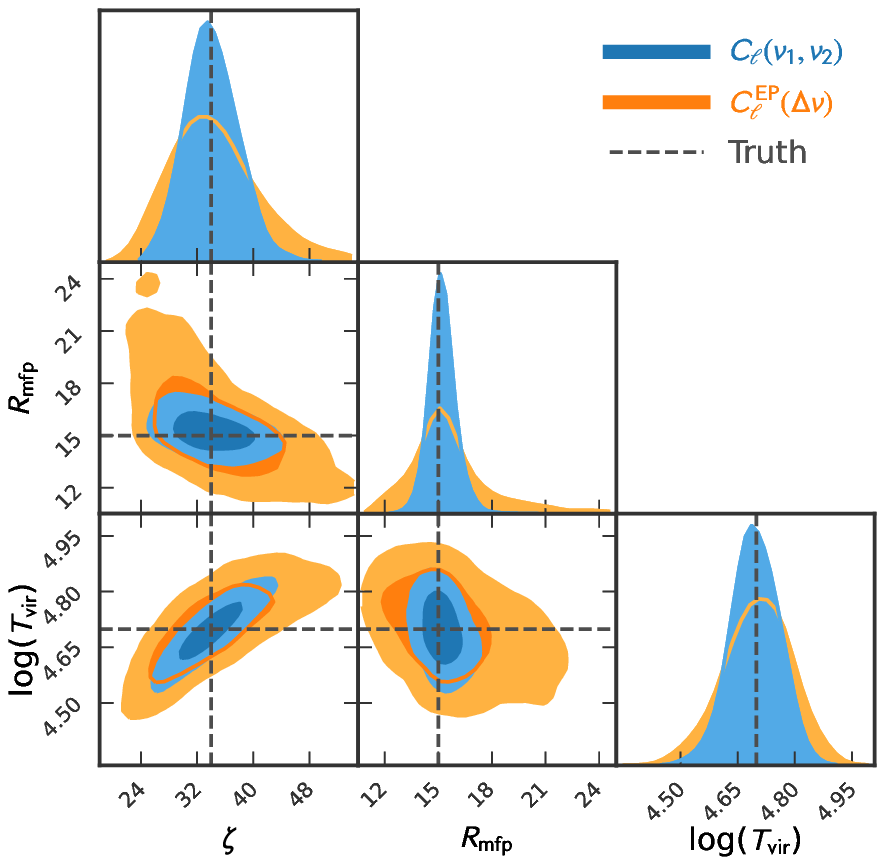}
\includegraphics[width=.33\textwidth, angle=0]{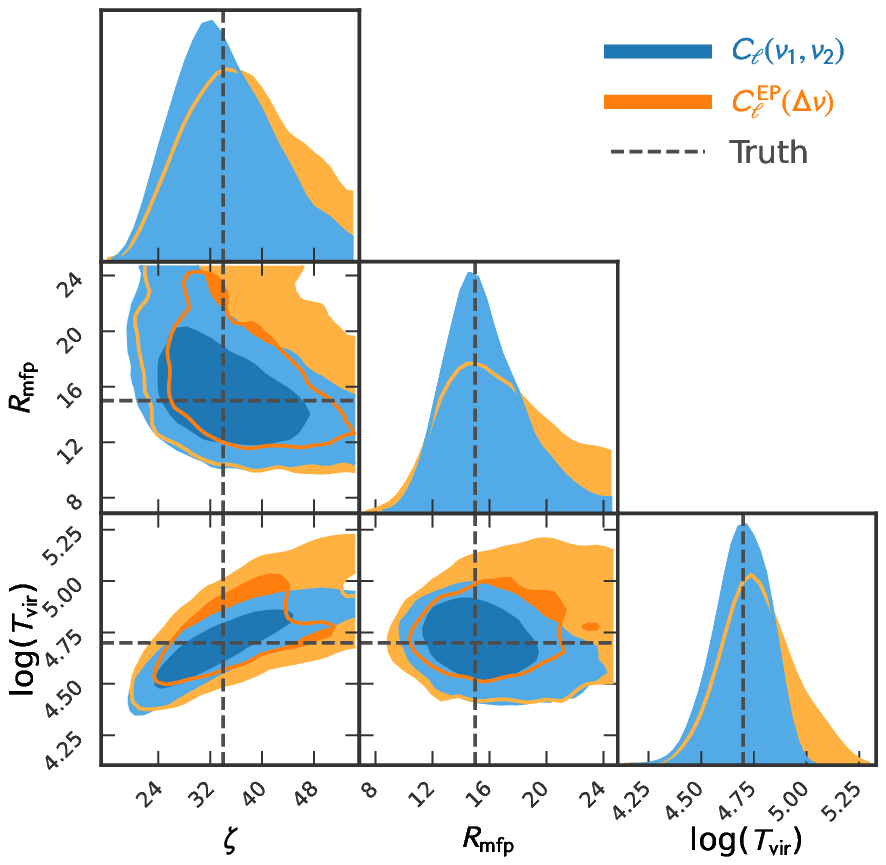}
\put(-422, 148){\Large {[a]}}
\put(-252, 148){\Large {[b]}}
\put(-82, 148){\Large {[c]}}
%
%
\caption{The posterior distribution (1D and 2D) of the parameters for [a] CV only (or $t_{\rm obs} = \infty$) [b] $t_{\rm obs} = 128$\,hrs [c] $t_{\rm obs} = 16$\,hrs, for MAPS (cyan) and power spectrum (orange). In the 2d marginalised posterior plots, the darker and lighter shading enclose 1$\sigma$ and 2$\sigma$ of the credible limits, respectively. The dashed lines denote the fiducial values of the parameters.}
\label{fig:corner}
\end{figure*}


\subsection{The MAPS emulator}
We develop a MAPS emulator based on artificial neural networks and use this as part of our MCMC analysis (see e.g. \citealt{Kern2017, Jennings2019}). This approach is very efficient and computationally fast. We use Latin Hypercube sampling to choose the parameter values for creating our set of training and testing {\sf 21cmFAST} simulation results. We have generated 1200 light-cone simulations with the same initial conditions in the parameter range $15 < \zeta < 55$, $5 < R_{\rm mfp} < 25$ and $4.0 < \log(T_{\rm vir}) < 5.4$, out of which 1100 were used for training and 100 were used for testing. Our network was built using the {\sf Keras}\footnote{\url{https://keras.io}} package, which runs on top of {\sf Tensorflow}\footnote{\url{https://www.tensorflow.org}}. The network consists of four hidden layers of sizes 16, 64, 256, and 1024, respectively. We refer the reader to \citet{Fling:2019} for a detailed description of the methodology. The accuracy of the emulator is quantified through the mean-square error which we find to be around 8 per cent.

\section{The error estimates}
\label{sec:error}
We consider two sources of errors in the measurements, namely instrumental noise and cosmic variance\,(CV). The system noise $\sigma_\ell^{\rm N}$ is an independent Gaussian random variable and it dominates the total error at small scales. At large scales the CV error $\sigma_\ell^{\rm CV}$ dominates. It arises from the finite volume of the universe accessible to the measurement and as shown by \citet{mondal15,mondal16}, the highly non-Gaussian nature of the 21-cm signal affects the CV error. However, it is unfortunately not computationally feasible to consider non-Gaussian effects in our CV error estimates. We therefore ignore the effects of non-Gaussianities. As our analysis is mostly sensitive to large scales where the non-Gaussianities are smaller, this is a reasonable assumption.

With regard to the instrumental noise, we consider three different scenarios in the context of a future SKA-Low observation. These are [a]\,a CV only case that corresponds to observation time $t_{\rm obs} = \infty$, [b]\,a deep survey with $t_{\rm obs} = 128$\,hrs, and [c]\,a shallow survey with $t_{\rm obs} = 16$\,hrs. The noise level of the latter is roughly equivalent to what can be achieved with $\sim 1000$ hours with LOFAR or MWA. Note that the entire analysis assumes the foregrounds are perfectly removed.

To reduce our data size as well as the noise we average the signals over 8 consecutive channels, which corresponds to $\sim 1.6$\,MHz. This is reasonable as we have found that the mean, as well as the statistics of the signal, do not change much over that frequency range. To calculate the combined effect of the CV and instrumental noise errors on the MAPS we use the same formalism as in \citet{mondal2020}, specifically their eq.\,13.

To account for the modelling error from the emulator as well as to suppress the contribution from small scales (as discussed in Sec.\,\ref{sec:method}), we introduce an additional 20\% modelling error i.e. $\sigma_\ell^{\rm m} = 0.2 \times\text{fiducial model MAPS}$. \citet{Mondal2020a}, who used an analogous emulator, employed a similar uncertainty in modelling. For a more detailed discussion, we refer the reader to the appendix of that paper. Thus, the total error in our mock data is given by $[\sigma_\ell^{\rm Tot}]^2 = {[\sigma_\ell^{\rm CV}]^2 + [\sigma_\ell^{\rm N}]^2 + [\sigma_\ell^{\rm m}]^2}$.

\section{Results}
\label{sec:result}
We use the publicly available {\sf emcee} package \citep{FM2013} to perform our MCMC parameter sampling using models from the emulator for each walker at each step in the chain. We assume uniform priors on all three parameters in the ranges: $15 < \zeta < 55$, $5 < R_{\rm mfp} < 25$ and $4.0 < \log(T_{\rm vir}) < 5.4$. We show the posterior distribution (1D and 2D) of the parameters in Fig.\,\ref{fig:corner} for the three different observational scenarios considered in this work. The corresponding best-fit values of the marginalised posterior and the $+/-$ 68\% upper/lower credible limits for each parameter are tabulated in Table\,\ref{tab:error}. The dashed lines in Fig.\,\ref{fig:corner} denote the fiducial values of the parameters. The results for MAPS are shown with a cyan colour and for SAPS with orange colour. A detailed description of the behaviour of the parameter posteriors can be found in \citet{Greig2015}. In this work, we are interested in the difference in constraints between using the MAPS and the SAPS as the metric for parameter estimation. From Fig.\,\ref{fig:corner}, we can see the parameters constraints are more stringent for the MAPS as compared to the SAPS. These improvements in constraining parameters come from the fact that the MAPS approach makes considerably better use of the available information from the 21-cm signal.

We would like to note that tests have shown that increasing the observing time beyond 128 hrs does not further reduce the credible limits on the parameters. This was found to be due to the assumed level of modelling errors. As modelling errors preferentially reduce the contribution of power from small scales, where the contribution of system noise errors in the total error budget is most important, we are basically not sensitive to these scales.


\begin{table}
\centering
\caption{The best-fit values and the $+/-$ 68\% upper/lower credible limits for each parameter.}
\label{tab:error}
\begin{tabular}{llll}

\hline\vspace{-.2cm} \\

{} & {} & $\dl(\nu_1, \nu_2)$ & $\dl^{\rm EP}(\Delta \nu)$ \\[4pt]

\hline\vspace{-.2cm} \\

{} & {CV only} & $33.693^{2.309}_{-2.037}$ & $33.846^{4.124}_{-3.744}$\\[4pt]
{$\zeta$} & $128$\,hrs & $33.862^{3.954}_{-3.414}$ & $33.846^{6.936}_{-5.536}$\\[4pt]
{} & $16$\,hrs & $33.710^{8.800}_{-6.693}$ & $36.915^{9.941}_{-8.332}$\\[4pt]

\hline\vspace{-.2cm} \\

{} & {CV only} & $15.218^{0.288}_{-0.272}$ & $15.140^{0.761}_{-0.600}$\\[4pt]
{$R_{\rm mfp}$} & $128$\,hrs & $15.167^{0.679}_{-0.580}$ & $15.243^{2.197}_{-1.524}$\\[4pt]
{} & $16$\,hrs & $15.366^{3.174}_{-2.288}$ & $16.524^{4.820}_{-3.533}$\\[4pt]

\hline\vspace{-.2cm} \\

{} & {CV only} & $4.693^{0.037}_{-0.039}$ & $4.703^{0.068}_{-0.069}$\\[4pt]
{$\log(T_{\rm vir})$} & $128$\,hrs & $4.695^{0.064}_{-0.060}$ & $4.705^{0.084}_{-0.090}$\\[4pt]
{} & $16$\,hrs & $4.699^{0.120}_{-0.130}$ & $4.759^{0.184}_{-0.151}$\\[4pt]

\hline
\end{tabular}
\end{table}


\section{Discussion and Conclusions}
\label{sec:conc}
We have compared the performance of two different statistical metrics for parameter estimation from mock light-cone observations of the 21-cm signal from the EoR. The first is the usual power spectrum $P(k)$, or more correctly the SAPS and the second is the MAPS $\cl(\nu_1, \nu_2)$. We find that the latter yields parameter constraints which have a factor $\sim 2$ smaller uncertainty,  both with and without the inclusion of instrumental noise. For our assumption on modelling errors, this advantage is already present for observation time 128\,hrs with the future SKA-Low.

As the MAPS metric extracts more information from the data than the SAPS metric, it is perhaps not so surprising that it performs better. The main problem is the invalid assumptions of periodicity and ergodicity along the LoS implied by the use of SAPS. Our results show that the extra information is not only relevant for parameter estimation but also extractable in the presence of realistic instrument noise. It would therefore seem advisable to base future work on parameter extraction on MAPS rather than SAPS.

The results in this letter constitute only a first exploration. We only consider one model and the improvement seen may be different for other models. We expect MAPS to work better the stronger the light-cone effect as it does not mix frequency and angular information the way SAPS does \citep[see e.g.][]{mondal18}. For the chosen model the average ionized fraction over the $\sim$29\,MHz light-cone changes by 0.42 which is substantial but not extreme. The significance of our results will decrease (increase) if a slower (faster) reionization history or shorter (wider) bandwidth is considered. We plan to quantify the improvement of MAPS over SAPS for different astrophysical parameters in future work. As part of this, we also plan to eliminate the emulator for the MAPS calculation and instead calculate it directly from the {\sf 21cmFAST} results. In that case it could be possible to decrease the modelling error used in the analysis. Furthermore, the calculation can possibly be made more efficient by neglecting those modes which lie furthest away from the diagonal and which likely contribute little in terms of information. Depending on the field of view which is being considered, it could also be good to take into account the curvature of the sky.

However, the largest challenge will be to gauge the impact of the (residual) foregrounds. The MAPS does not allow for foreground avoidance, which is possible for SAPS \citep{2010ApJ...724..526D, 2012ApJ...757..101T} and so it will be necessary to subtract the foreground signals. This will inevitably leave some residuals which may be strongly correlated over large distances in frequency and thus affect the MAPS. One advantage of using MAPS will actually be that any anomalous correlations between different frequency channels will be very obvious but the disadvantage is that they will interfere with the parameter estimation. However, an assessment of their actual impact requires a careful investigation based on actual foreground subtraction techniques and realistic residuals. It may be possible to perform an eigenmode analysis to eliminate the modes most affected by foregrounds residuals (see e.g. Koshambi-Karhunen-Lo\'eve eigenmode analysis in \citealt{Liu2012, Shaw2014}). Recent efforts to estimate MAPS in the presence of foregrounds show encouraging results \citep[see e.g.][and references therein]{2021MNRAS.501.3378P}. A first step could also be to obtain the actual MAPS data from currently existing data, such as LOFAR, HERA and MWA. Even though those signals are noise dominated, their properties can still be useful to increase our understanding of this promising summary statistic.

\section*{Acknowledgements}
This work is supported by the Wenner-Gren Postdoctoral Fellowship and by Swedish Research Council grant 	
2020-04691. RM would like to thank Prof. Somnath Bharadwaj for his useful discussion on MAPS. RM would also like to thank Charlie Fling and Prof. Ilian T. Iliev for the discussion on ANN. We have used the \textsf{pygtc} \citep{Bocquet2016} package for the corner plot. Parts of this research were supported by the Australian Research Council Centre of Excellence for All Sky Astrophysics in 3 Dimensions (ASTRO 3D), through project number CE170100013.

\section*{Data availability}
The data underlying this article will be shared on a reasonable request to the corresponding author.

\bibliographystyle{mnras} 
\bibliography{refs}


\vfill
\bsp
\label{lastpage}
\end{document}